\documentclass{vgtc}                          

\ifpdf
  \pdfoutput=1\relax                   
  \usepackage{graphicx}                
  \DeclareGraphicsExtensions{.pdf,.png,.jpg,.jpeg} 
\else
  \usepackage{graphicx}                
  \DeclareGraphicsExtensions{.eps}     
\fi%

\graphicspath{{figures/}{pictures/}{images/}{./}} 

\usepackage{microtype}                 
\PassOptionsToPackage{warn}{textcomp}  
\usepackage{textcomp}                  
\usepackage{mathptmx}                  
\usepackage{times}                     
\usepackage{cite}                      
\usepackage{tabu}                      
\usepackage{booktabs}                  
\usepackage[graphicx]{realboxes}
\usepackage{makecell}
\usepackage{soul}
\onlineid{0}

\vgtccategory{Research}

\title{Evoking empathy with visually impaired people through an augmented reality embodiment experience}

\author{Renan Guarese 
\thanks{e-mail: renan.guarese@rmit.edu.au}\\ %
        \scriptsize RMIT University %
\and Emma Pretty 
\thanks{e-mail: emma.pretty@rmit.edu.au}\\  %
     \scriptsize RMIT University  %
\and Haytham Fayek 
\thanks{e-mail: haytham.fayek@ieee.org }\\ %
     \scriptsize RMIT University %
 \and Fabio Zambetta 
 \thanks{e-mail: fabio.zambetta@rmit.edu.au}\\ %
     \scriptsize RMIT University
     \vspace{10pt}
\and Ron van Schyndel 
\thanks{e-mail: ron.vanschyndel@rmit.edu.au}\\ %
     \scriptsize RMIT University }

\teaser{
  \centering
\includegraphics[width=0.252\linewidth]{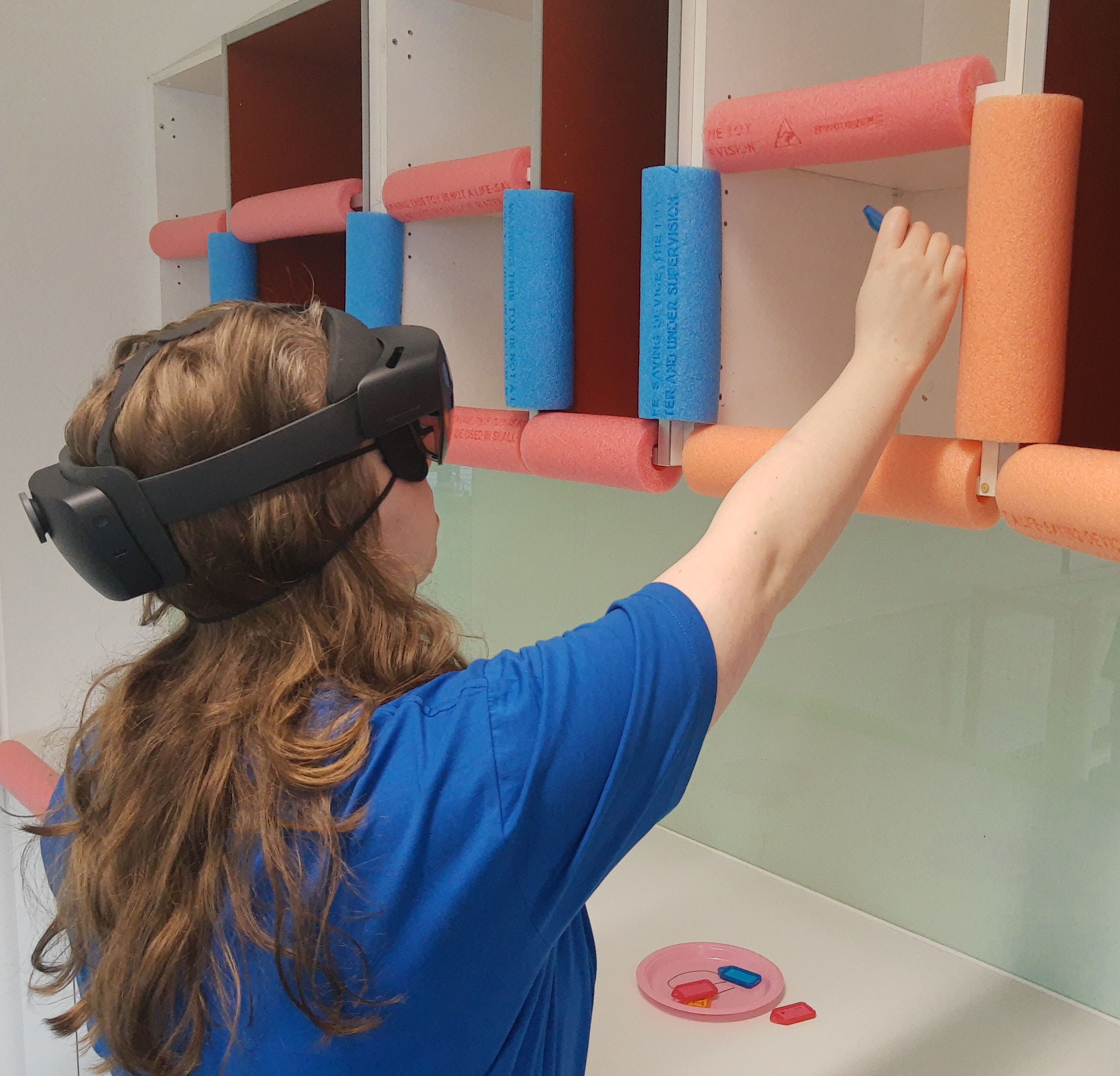}
\includegraphics[width=0.49\linewidth]{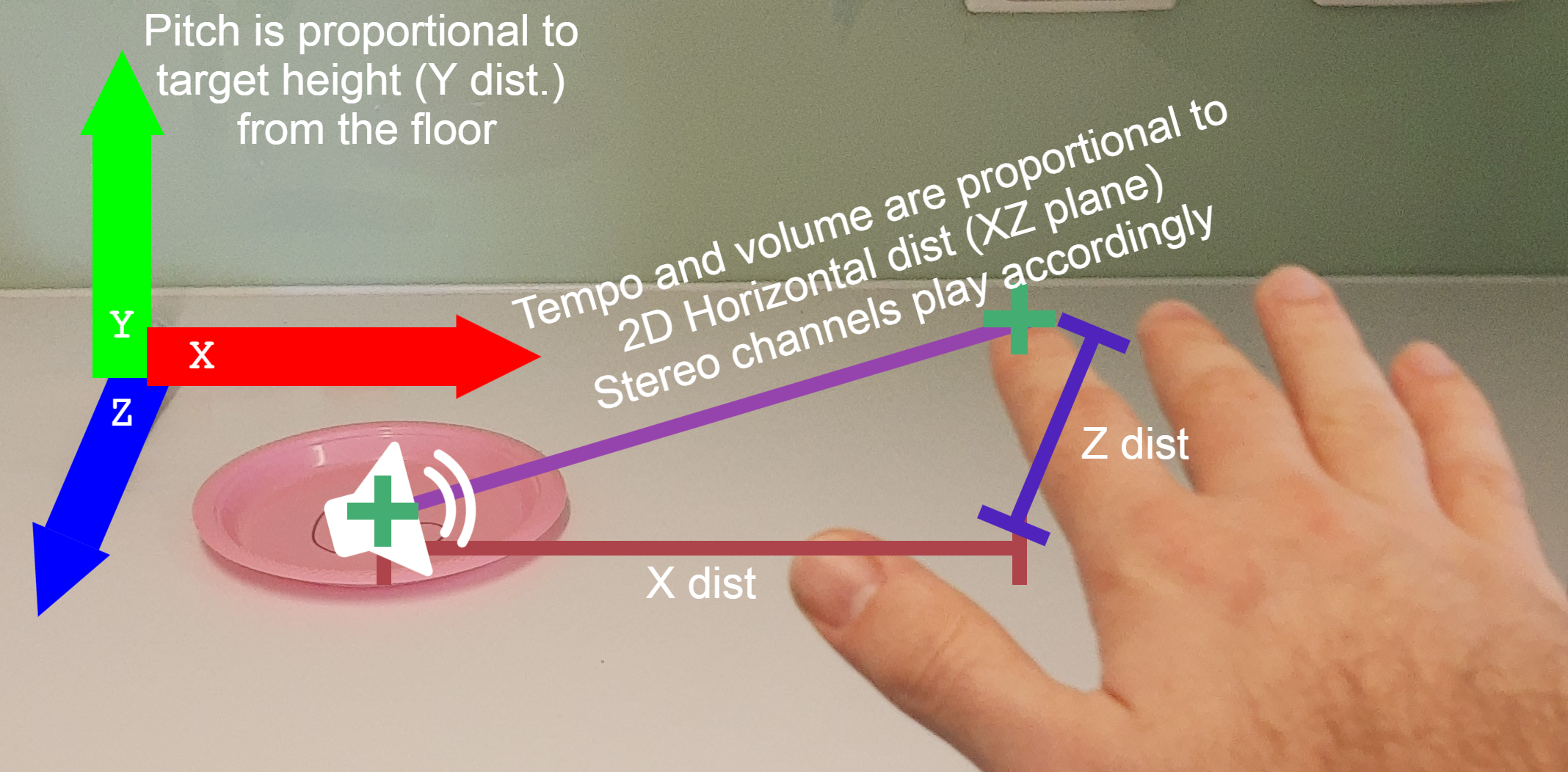}
\includegraphics[width=0.239\linewidth]{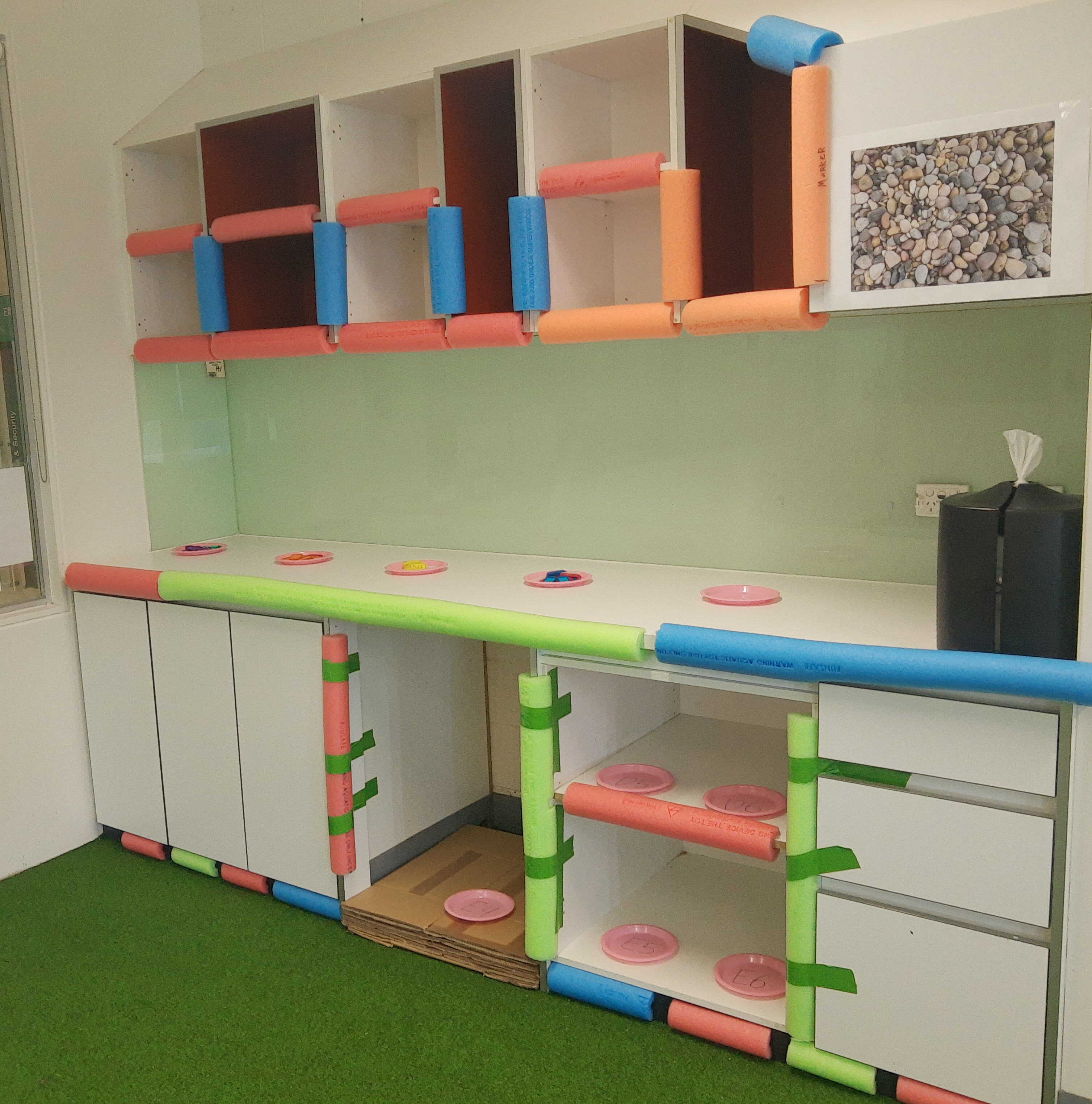}

  \caption{Left: a blindfolded user searching for a plate in a real cupboard and receiving virtual audio feedback from an AR device, during the experiment. Center: representation of the assistive technology used, with situated audio guidance coming from a target location, under a dynamic tempo, pitch, and stereo settings, varying according to the user hand position. Right: environment in which the experiment takes place - a large cupboard padded to avoid injuries -, along with physical objects for an increased sense of presence.}
  \label{fig:teaser}
}

\abstract{

To promote empathy with people that have disabilities, we propose a multi-sensory interactive experience that allows sighted users to embody having a visual impairment whilst using assistive technologies. The experiment involves blindfolded sighted participants interacting with a variety of sonification methods in order to locate targets and place objects in a real kitchen environment. Prior to the tests, we enquired about the perceived benefits of increasing said empathy from the blind and visually impaired (BVI) community. To test empathy, we adapted an Empathy and Sympathy Response scale to gather sighted people’s self-reported and perceived empathy with the BVI community from both sighted (N = 77) and BVI people (N = 20) respectively. We re-tested sighted people's empathy after the experiment and found that their empathetic and sympathetic responses (N = 15) significantly increased. Furthermore, survey results suggest that the BVI community believes the use of these empathy-evoking embodied experiences may lead to the development of new assistive technologies.} 

\CCScatlist{
  \CCScatTwelve{Human-centered computing}{Accessibility}{Accessibility technologies}{};
  \CCScatTwelve{Human-centered computing}{Accessibility}{Empirical studies in accessibility}{};
  \CCScatTwelve{Human-centered computing}{Human computer interaction (HCI)}{Interaction techniques}{Auditory feedback};
  \CCScatTwelve{Human-centered computing}{Human computer interaction (HCI)}{Interaction paradigms}{Mixed / augmented reality};
}

\begin{document}



\maketitle


\section{Introduction}

Digital technologies have emerged in recent years as an avenue for encouraging social change and personal growth~\cite{empathyLitReview}. In particular, technologies such as virtual and augmented reality (VR and AR), as well as video games, are immersive and widespread communication channels~\cite{sherman2018understanding}. Common applications of video games for social change, also known as serious games, are educating users about the impact of climate change on society and the environment~\cite{flood2018adaptive}, 
and the gamification of healthy behaviours to encourage regular physical activity and the consumption of a healthy diet~\cite{ma2014future}. 

VR and AR experiences for social change have made use of faithful sensory simulation to place users in an environment or scenario that they wish to bring awareness to~\cite{kandaurova2019effects}. These applications have also utilised the interactions between the physical world and virtual projections of information to simulate embodiment of the experiences from other people, such as the embodiment of one user's emotions to the other in real-time~\cite{semertzidis2020neo}, the experiences of being of a different gender~\cite{yee2011men}, nationality~\cite{groom2009influence}, physical stature~\cite{yee2007proteus}, or having a disability~\cite{bacchus2019life, calepso2020, Guarese:Xchange, guareseEmpathy:2021}. These experiences are examples of `embodiment of difference', which are usually avoided by people not in underrepresented or minority groups~\cite{riley2013island}. However, such technologies can be instrumental in evoking empathy or understanding to create social change and improving the lives of people with a disability~\cite{empathyLitReview}.

Evoking empathy has proven to be effective in encouraging the creation of accessible technologies and applications that consider the diverse needs of its users~\cite{raviselvam2021simulation}. For example, a workshop which involved participants experiencing simulated visual impairment was effective in both evoking empathy with blind and visually impaired (BVI) people and enhancing the novelty and creativity of proposed design solutions to better guide those with visual impairments~\cite{raviselvam2016user}. Empathy is also a valuable tool in reducing stigma towards disabilities and mental illnesses~\cite{chowdhury2019vr, ando2011simulation, penn2010virtual}, which leads to an overall improvement to the lives of all affected people.

We posit embodiment of experiences as a way to encourage empathy with BVI people, reduce stigma, and stimulate the design and production of assistive technologies. Previous research has claimed that the responsibility for addressing disability should be placed on the collective, including technology designers~\cite{Thieme2018, Branham:2015}, and we believe that by increasing empathy with those that have a disability, this responsibility is more likely to translate to action. 

Whilst the intention of this research is to bring awareness and kindness towards the BVI, efforts to evoke empathy can prove detrimental and the design of the embodiment of difference experience can play a role in the outcomes. Prior work~\cite{silverman:2015} found that participants engaged in a challenging blindness simulation ended up judging blind people as less capable of performing work and independent living. However, the same researchers agree that simulating other people’s experiences improves attitudes toward those people. By exploring this balance in which the gains in sympathy and empathy with people who have disabilities outweigh the risks of patronizing that same demographic, we believe there can still be a positive impact in the development of next-generation assistive technologies. Thus, instead of allowing people without disabilities to solely perform brief experience simulations of disabilities, we aim to engage them in simulating new assistive technology methods while having their vision completely occluded, in an attempt to replicate a total loss of vision. 


This work contributes a user study of an embodiment of difference experience using Audio AR \cite{yang2022audio}. To our knowledge, it is the first demonstration of an empathy-evoking AR experience that totally occludes the vision of sighted people whilst guiding them with a proposed assistive technology. Instead of demonstrating this concept in a purely virtual environment (VR), being blindfolded in AR allows sighted users to have access to dependable and accessible guidance \textit{in loco}, since the real environment can be mapped by the device at runtime. This way, users can roam a real environment and touch real objects while receiving audio guidance, having the exact same experience a totally blind person would have in sonified AR. Thus, we add AR to the body of literature that states that VR is a successful tool for evoking empathy that aims to create social change by improving the perceptions and raising awareness for those that are different~\cite{empathymachine2020}. It also uniquely involves the input and advisement of a visually impaired researcher (one of this paper's co-authors) in an attempt to authentically create an experience that explores the different senses BVI people rely on, and how they can make use of assistive technologies. As such, a regular blindfold is used to occlude any real or virtual visual aspects, as depicted in Fig. \ref{fig:teaser}(left). An AR headset was used to provide the user with spatialized audio, which has been shown to increase user immersion \cite{potter:2022}, augmenting a kitchen space to be specially accessible for BVI people, with spatial audio guidance cues to find objects, similar to~\cite{Iravantchi:2020, May:2019, guarese:2021}. 

This work also provides, for the first time, a comparison of the perceived and self-reported empathy of sighted people with the BVI community to gain insight into the differences between the BVI community's experience with sighted people and sighted people's feelings towards the BVI community. We hope to encourage the inclusion of AR devices in digital empathy research, as well as the use of surveys that explore the discrepancies between the BVI community and those engaging with the embodiment of difference experiences to gain a better understanding of said discrepancy. Further, in an effort to validate this research, a survey was provided to the BVI community (N = 20) with questions regarding their opinions on the need of empathy-evoking experiences, and whether they believe these experiences to be beneficial. 

The remainder of this paper is organised as follows: Section~\ref{section:relatedworks} provides an overview of works in embodiment of difference in general, and how blindness has been approached in AR and VR. Subsequently, Section~\ref{section:hypotheses} present the hypotheses that motivate this study, and Section~\ref{section:methodology} defines the methodology used to assess them, divided into a preliminary survey of the community, the experiment design, the audio interface and the metrics used in said experiment, and finally the feedback from an expert who tested it. Section~\ref{section:results} details the findings from our work, followed by a discussion about their meanings (Section ~\ref{section:discussion}), before concluding and making suggestions for future work (Section ~\ref{section:conclusion}).

\section{Related Works}
\label{section:relatedworks}

We begin by exploring how virtual and augmented environments can support the embodiment of difference, how blindness has been approached in VR and AR simulations, and in what forms has digital empathy been studied in the HCI community \cite{empathyLitReview}.

\subsection{Embodiment of difference} Mixed and virtual environments have the capability to supply users with sensory cues that can empower experiences with comparable features~\cite{ahn2011embodied, guareseEmpathy:2021}. Recently, self-representation in virtual reality has been explored and applied in several works in order for users to perceive experiencing disabilities and conditions from other people's perspectives~\cite{bacchus2019life, calepso2020, guareseEmpathy:2021, kandaurova2019effects, bujic2020empathy}. For example, a study from 2013~\cite{peck_black} utilised VR to embody white people in bodies of people with white, purple, and dark skin. Pre and post-measurements of implicit racial bias, that is the linking of negative words with people with dark and positive words with people of white skin, found that implicit racial bias significantly decreased after spending approximately 12 minutes in the bodies of dark skinned people.

In a 2019 empathy experience idealised by expressionist artist Marcel Schreur~\cite{bacchus2019life}, a unique life perspective is explored in VR. Schreur's life as a thirty-year oral cancer and seven-year vascular dementia survivor is depicted with his struggles, attitude towards life and singular viewpoint being explorable in first-person. In a similar manner, a 2020 digital empathy project brought to VR an amputee's embodiment experience~\cite{calepso2020, Guarese:Xchange}, allowing users to feel a virtual rendering of the Phantom Limb Sensation. These works generate a space that allows participants to consider aspects of their own mental and physical processes through a lens of difference and disability. In~\cite{guareseEmpathy:2021}, researchers proposed an experience with the same ambition of providing empathy with difference, however focusing on blindness. In order to assess their proposal that sighted people's views regarding BVI people’s experiences might be affected by interacting with their game, we have created a similar experience simulating the use of assistive technologies for guidance while being blindfolded. As to properly measure that effect, a standardized questionnaire~\cite{Escalas:2003} was adapted to gauge their previous and post levels of sympathy and empathy with that demographic. 

Other research has also adapted the sympathy and empathy questionnaire in~\cite{Escalas:2003} for immersive VR experiences, though the exact adaptions of each question were not provided. In~\cite{kandaurova2019effects}, a comparison of viewing a non-interactive 3D video about social issues around the globe to a 360 degree VR replica of the same video found increased empathy and a higher intention to donate their time or volunteer for the social cause when watching the VR experience. Similarly,~\cite{bujic2020empathy} compared a 360 degree VR video, the same video but in 2D, and a written article on the subject of refugee crisis in Syria. The VR video was found to increase empathy the most, and to be the most likely to elicit a positive change in attitude.


\begin{table*}[t]
\caption{Survey provided to the community prior to the experiment, including the mean (M) and standard deviation (ST) of the results.}

\label{surveyTable}
\begin{tabular}{l|l|l}

  \thead{ 1 | 2 | 3 | 4 | 5 | 6 | 7 \\ \includegraphics[width=0.22\linewidth]{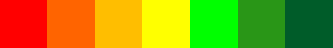} \\ \textbf{ Likert Scale Statements}  (1 = "Completely \textbf{Disagree}"  7 = "Completely \textbf{Agree}")}
 
 & \thead{BVI Respondents (N = 20)} & \thead{Sighted Respondents (N = 77)}  \\ \hline
\thead{I feel as though there is a need for sighted people to better understand what BVI \\people go through on their day-to-day tasks.} 
& \thead{M = 5.8  ST = 1.79 \\ 1 \includegraphics[width=0.18\linewidth]{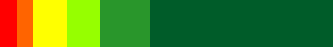} 7
}  
&  \thead{M = 6.17  ST = 1.29\\ 1 \includegraphics[width=0.18\linewidth]{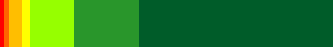} 7}    \\ \hline

\thead{I believe that blindfolding a sighted person is enough for them to create empathy \\towards BVI people.} 
& \thead{M = 3.00  ST = 2.02 \\ 1 \includegraphics[width=0.18\linewidth]{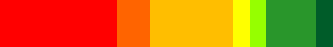} 7}  

&  \thead{M = 3.13  ST = 1.56 \\ 1 \includegraphics[width=0.18\linewidth]{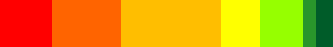} 7} \\ \hline

\thead{I believe that the different BVI levels and conditions need to be demonstrated to \\sighted people in order for them to develop empathy towards BVI people.   }
& \thead{M = 4.8  ST = 1.93\\ 1 \includegraphics[width=0.18\linewidth]{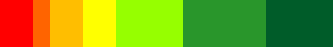} 7}    
&  \thead{M = 5.54  ST = 1.58 \\ 1 \includegraphics[width=0.18\linewidth]{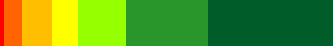} 7} \\ \hline

\thead{On top of demonstrating how a BVI person perceives the world (e.g., blindfolding), \\I believe sighted people interacting  with assistive technologies for BVI people \\ (e.g., a screen reader or voice commands) can also improve their empathy.}
& \thead{M = 4.9  ST = 2.26\\ 1 \includegraphics[width=0.18\linewidth]{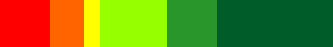} 7}    
&  \thead{M = 5.45  ST =  1.42  \\1 \includegraphics[width=0.18\linewidth]{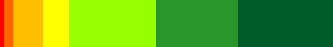} 7} \\ \hline

\thead{I believe that by improving sighted people's empathy towards BVI people, \\awareness about the needs and experiences of BVI people can be raised.  }
& \thead{M = 5.15  ST = 2.00\\ 1 \includegraphics[width=0.18\linewidth]{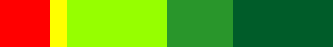} 7}    
&  \thead{M = 6.24  ST = 1.00  \\ 1 \includegraphics[width=0.18\linewidth]{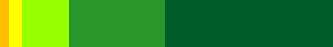} 7} \\ \hline

\thead{I believe that by improving sighted people's empathy towards BVI people through \\simulated experiences, new technologies will emerge that assist these people.}
& \thead{M = 4.75  ST = 2.08\\ 1 \includegraphics[width=0.18\linewidth]{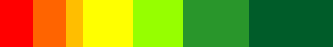} 7}    
&  \thead{M = 6.05  ST = 1.24  \\1 \includegraphics[width=0.18\linewidth]{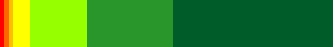} 7} \\ 

\end{tabular}
\end{table*}

\subsection{Blindness in VR and AR}
Recently, both commercial and academic settings have seen the implementation of blindness simulation experiences in mixed and virtual reality. The 2016 immersive documentary \textit{Notes on Blindness}\footnote{http://www.notesonblindness.co.uk/vr/} was released alongside a homonymous feature film, based on a man's sensory and psychological experience with his own loss of vision. The events are composed of non-interactive 360-degree videos of visual elements representing where spatial sounds are coming from, such as the silhouette of a person walking. Similarly, a 2018 VR game named \textit{Blind}\footnote{https://www.tinybullstudios.com/games/blind/} also conveys blindness by rendering visual details of spaces based on echolocation, accompanying  players throughout their paths as a blind character that needs to solve several puzzles. In the same fashion of the commercial ones, a 2020 study~\cite{vrcontest2} used visual cues to aid users in guiding themselves with a virtual walking cane. Sounds propelled from cane collisions, footsteps and characters' chatter create light elements, forming a mesh of the environment. Other works have studied AR/VR simulation of specific visual impairment conditions regarding their visual accuracy, such as glaucoma \cite{jones2020} or cataracts \cite{krosl:2020}, which fall out of the scope of the current study. Unlike these, our experiment aims to completely occlude any visual elements, focusing on a total blindness condition, supplying users solely with spatial audio cues and their own sense of touch. These are meant to feel much like a newly blinded person would, prior to adjusting to their loss of vision. 

Another 2020 VR study~\cite{vrcontest1} executed similar navigation tasks without visual cues. Their audio guidance, however, is mainly comprised of audio description of the path and environment, with an AI agent continuously mentioning the correct direction and any possible obstacles to be avoided. A beeping sound also warns the users when they get close to obstacles, similar to the collision warning systems of modern cars, however these do not seem to be spatialized. In contrast to these works and previous embodiment of difference experiences, the current application focuses on providing the user with an AR approach, being set on an actual kitchen environment, with passive haptics (i.e. allowing people to touch the real objects), spatial audio localization cues and audio instructions. Unlike in a purely virtual environment (VR), we believe AR offers users the ability to roam and touch the real environment around them, while receiving audio guidance for its objects, having the exact same experience a totally blind person would have in sonified AR.

\section{Hypotheses}
\label{section:hypotheses}

We believe that there is a large disconnect between sighted people’s empathy with the BVI community and BVI people's perception of said empathy. In the context of promoting a sense of empathy with the difference in people without disabilities, our AR experience enables sighted users to embody having a visual impairment while relying on a proposed assistive technology for day-to-day tasks. However, in order to assess whether their empathy with BVI people is significantly affected by the experiment, it is necessary to objectively quantify that response and compare its value before and after the experience. Thus, a sympathy and empathy measurement tool~\cite{Escalas:2003} will be applied to sighted users before and after the experiments. The goal is to understand whether and how their previous views regarding BVI people’s experiences changes after interacting with the experiment, with the intention of raising awareness to the need for assistive technologies aimed at visual disabilities. Thus, the following hypotheses were the basis of this research:

\smallskip
\smallskip

\textbf{\textit{H1: There is a significant difference between sighted people’s empathy with the BVI community and BVI people's perception of sighted people’s empathy with the BVI community.}}


\smallskip
\smallskip

\textbf{\textit{H2: Sighted people’s empathy with the BVI will increase after experiencing a proposed assistive technology using sonified AR for micro-guidance tasks.}}

\section{Methodology}
\label{section:methodology}
In respect to the motto "Nothing about us without us"~\cite{charlton1998nothing}, the work done here is partly authored by a visually impaired researcher with advice and opinions sought from the BVI community via an internet survey. Additionally, it included the feedback of an external visually impaired researcher, who was invited to participate as an expert in digital accessibility, providing valuable insights about the sonified AR experience, prior to the start of user tests.

\subsection{Preliminary Survey}

In an effort to validate this research in the first place, a survey was provided to the BVI community (N = 20) with questions regarding their opinions on the need of empathy-evoking experiences, and whether they believe these experiences to be beneficial. For instance, when asked if they felt there was a need for sighted people to better understand what they go through on their day-to-day, 16 people (80.00\%) agreed to it. 14 respondents (70.00\%) also believed that by having sighted people interact with assistive technologies for BVI people, the sighted can also improve their empathy, while 12 people (60.00\%) agreed that by improving their empathy, new assistive technologies will emerge. The remainder of the results can be seen in Table \ref{surveyTable}. Based on these results, we believe this work explores a valuable gap that is in line with the BVI community's beliefs and has a potential outcome not solely on increasing empathy across society, but also on the development of novel assistive technologies.

\subsection{Experiment scenario and tasks}
\label{experiment}

The experiment scenario portrays a space specially adapted to be accessible for BVI people, being augmented with audio guidance cues to find objects, similar to~\cite{Iravantchi:2020, May:2019, guarese:2021}. Particularly, a real kitchen with a large cupboard, counter and high shelves (Fig. \ref{fig:teaser}-right) is used, instead of a fully virtual one \cite{May:2019}. The guidance methods were implemented by using the Microsoft HoloLens 2\footnote{https://www.microsoft.com/hololens/} because of its spatialized audio system (using its standard HRTF settings\footnote{https://learn.microsoft.com/en-us/windows/mixed-reality/design/spatial-sound}), spatial mapping, and hand-tracking features. All user tasks require a sense of micro-guidance by the user, i.e. close-range navigation and object-handling tasks~\cite{guarese:2022}, which is provided by the scenario via a proposed assistive technology: spatialized audio guidance cues. Objects need to be placed in locations found based solely on sound and tactile guidance. The kitchen furniture, plates and tokens present in the experience are real physical objects. These are meant to improve the user's sense of presence during the experiment, embodying them with passive haptics~\cite{Insko:2001}, i.e. the texture, weight, and hardness of the real objects will be felt by the user. All surfaces and furniture edges that might have been harmful to users were padded with polyethylene foam, as to avoid any injuries, as can be seen in Fig. \ref{fig:teaser}-right. 

Blindfolded participants were provided audio instructions on how to follow each of the audio guidance cues, as well as test them during a learning phase. Afterwards, they performed 30 object-placement trials on plates at different positions in the cupboard, while following the audio guidance. The order of the trials was counter-balanced between users, following a Latin Square. After the tests, participants were asked to respond to a questionnaire regarding their experience.

\subsection{Proposed Assistive Technology used}

Visually impaired people mostly rely on their hearing and touch senses to orient themselves to their surroundings~\cite{lighthouse:2015}. However, some surfaces can be dangerous, uncomfortable, or even unsanitary to touch, such as a hot stove, or a soiled garbage can. By tracking and guiding the user's hand position via sound in AR, dynamic spatialized audio feedback can precisely inform the location of a kettle handle or the top of a bin \cite{guarese:2022}. In this manner, the user is able to appropriately place their hand to perform the task, without having to touch any undesired surfaces. 

Similar to recent studies~\cite{May:2019, guarese:2021, barde:2020, pare:2021}, an object-targeting auditory interface was used to render audio guidance cues to blindfolded users in the 3D space surrounding them. Based on their results, three audio components were used to convey guidance to targets located at different places in the 3D scenario, as depicted in Fig.\ref{fig:teaser}-center:

\begin{itemize}
        \item Tempo: The 2D horizontal distance (X and Z axes) from the user hand to the target is conveyed by an on-off alternation (tempo) of the sonification, always maintaining a constant “off” interval (0.03 s) and varying the length of its “on” signal to a sixth of the current distance (cm) in seconds, similar to \cite{guarese:2022}. 
        The closer the hand is to the target, the lesser will be the wave’s mark-to-space ratio, i.e., tempo will sound faster. In the same sense, the farthest it gets to the target, the wave’s mark-to-space ratio will be greater, i.e., its tempo will sound slower. This audio metaphor is similar to the beeping parking sensors of modern cars, which indicate proximity, and is also frequently used in sound interfaces to convey urgency~\cite{edworthy:1995}. 
        \item Stereo: Audio guidance is demonstrated in a spatialized fashion, similar to several works~\cite{pare:2021, guarese:2021, barde:2020}. Regardless of user orientation, if the target is positioned to the right of their hand, sound will only be heard in the right ear. Likewise, if the sound comes from the left ear, it means that the target is to the left of their hand, with the sound flipping from side to side as the hand crosses past the target horizontally. Additionally, the amplitude (volume) decreases in a linear roll-off as the hand distance to the target increases, and vice-versa.  
        \item Pitch: Results from several studies replicated a human notion that higher pitched sounds tend to be associated with more highly elevated objects, and vice-versa \cite{Evans:2010, Neuhoff:2002}. In that sense, a direct mapping between target height and pitch was used. The lowest pitch being played indicated the lowest possible target height (at floor level), with the highest pitch indicating the highest possible target height (top shelf level). Targets in-between these heights used in-between pitches, following a parabola, as to spread them apart equally in audio perception. Particularly in this experiment, as to account for audio usability~\cite{buzzScale}, the range of fixed pitches was defined by: a) starting at a low frequency that is high enough to be heard by most people, that does not present any virtual distortion in the sound equipment used, and that does not cause any noticeable discomfort to the ears (\textit{ad hoc} 230 Hz); b) ending at a high frequency that is low enough to abide by the same criteria (\textit{ad hoc} 1200 Hz); and c) having the final frequency values for each height to be represented by the closest musical note in pitch, without repeating the same note at different octaves in multiple heights, as to avoid bias. 
        
        
        By following these constraints, the notes attributed to each height of the cupboard were: 
        \begin{itemize}
        \item Top shelf: D6 (1175 Hz)
        \item Second-from-the-top shelf: A5 (880 Hz)
        \item Counter: C5 (523.3 Hz)
        \item Second-from-the-bottom shelf: E4 (329.6 Hz)
        \item Bottom shelf: B3 (246.9 Hz)
        \end{itemize}

\end{itemize}

\subsection{Empathy and Sympathy scales}

To assess if there is any change in emotional response from a user without disabilities towards the BVI population with the impairment being investigated, a measuring tool is required. As our hypotheses are specifically targeted at empathy, it is necessary to quantify this emotional response from users before and after the experiment takes place. To empirically test sympathy and empathy responses of spectators with characters depicted in advertising videos, reliable measurement instruments were developed by Escalas and Stern~\cite{Escalas:2003}: the Ad Response Sympathy (ARS) and Ad Response Empathy (ARE) scales. Their results indicated that sympathy responses mediated the effect of a drama advertisement's form on empathy responses, with both sympathy and empathy directly enhancing positive attitudes to an advertisement. Following this notion, the current work will abide by the same concepts of sympathy and empathy, by hypothesizing that these metrics can be applied to different forms of media. As the statements in their tools refer specifically to users watching a video, they were adapted into three differently phrased versions (as shown in Table \ref{scaleAdaptionTable}), each with a slight different purpose:


\clearpage

\begin{table*}[t]
\caption{Sympathy (ARS) and Empathy (ARE) statements from Escalas and Stern \cite{Escalas:2003}, and its according adapted versions.}
\label{scaleAdaptionTable}
\Rotatebox{90}{%
\begin{tabular}{|l|l|l|l|}
\hline
\textbf{Original statement\cite{Escalas:2003}} 
& \textbf{BVI perception survey} 
& \textbf{Sighted pre-test (previous experience)}
& \textbf{Sighted post-test (after experiment)} 
\\ \hline

Ad Response Sympathy (ARS) scale
& Adapted ARS I
& Adapted ARS II
& Adapted ARS III
\\ \hline

\thead{\textbf{ARS-1:} Based on what was happening \\in \hl{the commercial}, \\ I understood what \hl{the characters} \\\hl{were} feeling.} 
& \thead{Based on what was happening \\in \hl{my experiences}, \\I \hl{believe sighted people} understood what \hl{I} \\\hl{was} feeling \hl{in my day-to-day tasks.}} 
& \thead{Based on what was happening \\in \hl{my experiences}, \\I understood what \hl{BVI people} \\\hl{were} feeling \hl{in their day-to-day tasks.}}
& \thead{Based on what was happening \\in \hl{the experiment}, \\I understood what \hl{BVI people} \\\hl{are} feeling \hl{in their day-to-day tasks.}} 
\\ \hline 

\thead{\textbf{ARS-2:} Based on what was happening \\in \hl{the commercial}, I \\understood what was bothering \\\hl{the characters.}} 
& \thead{Based on what was happening \\in \hl{my experiences}, I \\\hl{believe sighted people} understood what \\was bothering \hl{me in my day-to-day tasks.}} 
& \thead{Based on what was happening \\in \hl{my experiences}, I \\understood what was bothering \\\hl{the BVI people in their day-to-day tasks.}} 
&  \thead{Based on what was happening \\in \hl{the experiment}, I \\understood what \hl{is} bothering \\\hl{BVI people in their day-to-day tasks.}} 
\\ \hline

\thead{\textbf{ARS-3:} While \hl{watching the ad,} \\I tried to understand \\the events as they occurred.} 
& \thead{While \hl{experiencing my day-to-day tasks,} \\I \hl{believe sighted people} tried to understand \\the events as they occurred.} 
& \thead{While \hl{perceiving the day-to-day tasks of} \\
\hl{BVI people,} I tried to understand \\the events as they occurred.} 
&  \thead{While \hl{experiencing the simulation,} 
\\I tried to understand the events as 
\\they \hl{were occurring to a BVI person.}} 
\\ \hline

\thead{\textbf{ARS-4:} While \hl{watching the ad,} \\I tried to understand the \\\hl{characters'} motivation.} 
& \thead{While \hl{experiencing my day-to-day tasks,} \\I \hl{believe sighted people} tried to understand \\the motivation \hl{of my behaviors.}} 
& \thead{While \hl{perceiving the day-to-day tasks of}\\
\hl{BVI people,} I tried to understand the \\motivation \hl{of their behaviours.}} 
&  \thead{While \hl{experiencing the simulation,} \\I tried to understand the motivation \\\hl{of BVI people in their day-to-day tasks.}} 
\\ \hline

\thead{\textbf{ARS-5:} I was able to recognize the \\problems that \hl{the characters in the ad} had.} 
& \thead{
  I \hl{believe sighted people were} able to recognize
  \\the problems that \hl{I} had \hl{in my experiences.}} 
& \thead{I was able to recognize the problems that \\\hl{BVI people} had \hl{in my experiences with them.}} 
&  \thead{I was able to recognize the problems that
  \\\hl{BVI people have in their day-to-day tasks.}} 
\\ \hline

Ad Response Empathy (ARE) scale
& Adapted ARE I
& Adapted ARE II
& Adapted ARE III
\\ \hline

\thead{\textbf{ARE-1:} While \hl{watching the ad,} \\I experienced \\feeling as if \\the events were really happening to me.} 
& \thead{While \hl{performing my day-to-day tasks,} \\I \hl{perceived sighted people to be}\\ \hl{experiencing} feeling as if the \\events were really happening to \hl{them.}} 
& \thead{While \hl{perceiving the day-to-day tasks of}\\
\hl{BVI people,} I experienced feeling as if \\the events were really happening to me.} 
&  \thead{While \hl{experiencing the simulation,} \\I experienced \\feeling as if \\\hl{I really had a visual impairment.}} 
\\ \hline

\thead{\textbf{ARE-2:} While \hl{watching the ad,} \\I felt as though \\I were one of the characters.} 
& \thead{While \hl{performing my day-to-day tasks,} \\I \hl{perceived that sighted people} felt as though \\\hl{they} were \hl{a BVI person.}} 
& \thead{While \hl{perceiving the day-to-day tasks}\\
\hl{of BVI people,} I felt as though \\I were \hl{a BVI person.}} 
&  \thead{While \hl{experiencing the simulation,} \\I felt as though \\I were \hl{a BVI person.}} 
\\ \hline

\thead{\textbf{ARE-3:} While \hl{watching the ad,} \\I felt as though \\\hl{the} events \hl{in the ad} \\were happening to me.}
& \thead{While \hl{performing my day-to-day tasks,} \\I \hl{perceived that  sighted people} felt as though \\\hl{those} events \\were happening to \hl{them.}}
& \thead{While \hl{perceiving the day-to-day tasks of}\\
\hl{BVI people,}  I felt as though \\\hl{those} events were happening to me.} 
&  \thead{While \hl{experiencing the simulation,} \\I felt as though \\\hl{I had a visual impairment.}} 
\\ \hline

\thead{\textbf{ARE-4:} While \hl{watching the commercial,} \\I experienced many of the same \\feelings that \hl{the characters portrayed.}} 
& \thead{While \hl{performing my day-to-day tasks,} \\I \hl{perceived that sighted people} experienced \\many of the same feelings that \hl{I expressed.}} 
& \thead{While \hl{perceiving the day-to-day tasks of}\\
\hl{BVI people,} I experienced many of the \\same feelings that \hl{they expressed.}} 
&  \thead{While \hl{experiencing the simulation,} \\I experienced many of the \\same feelings that \hl{BVI people express.}} 
\\ \hline

\thead{\textbf{ARE-5} While \hl{watching the commercial,} \\I felt as if \hl{the characters'} feelings \\were my own.} 
& \thead{While \hl{performing my day-to-day tasks,} \\I \hl{perceived that sighted people} felt as if \hl{my} \\feelings were \hl{their} own.} 
& \thead{While \hl{perceiving the day-to-day tasks of}\\
\hl{BVI people,} I felt as if \hl{their} feelings \\were my own.} 
&  \thead{While \hl{experiencing the simulation,} \\I felt as if \hl{BVI people's} feelings \\were my own.} 
\\ \hline

\end{tabular}
}%
\end{table*}

\clearpage

\begin{itemize}
        \item BVI perception survey: In order to evaluate \textbf{H1}, it is first necessary to understand how the BVI community perceives sighted people's empathy with them as individuals. This is important not only
        to check if, as hypothesized, there is a difference in empathy, but also to validate the need for an empathy-evoking experience. Thus, an adapted version of Escalas and Stern's scale \cite{Escalas:2003} was applied as an online survey to BVI people. The adaptation concerned their overall personal experiences with sighted people, rather than video dramas, and how they perceived sighted people felt towards them in several criteria. 

    \item Sighted pre-test: Similar to the aforementioned adaptation, by surveying sighted people on their overall past experiences with BVI people in an equivalent set of criteria, it is possible to compare their self-reported results to the BVI people’s perception. This comparison is vital to assess the validity of \textbf{H1}. Beyond that, a measurement of their empathetic and sympathetic responses towards BVI people prior to their participation in the experiment is required to determine whether or not there is a change, as \textbf{H2} suggests. Hence, another adaptation of the ARS and ARE scales \cite{Escalas:2003} was provided as an online survey to sighted people prior to the experiment taking place. This version, instead of mentioning video dramas, referred to their overall personal experiences with BVI people and how they personally felt. 
    \item Sighted post-test: Finally, as to gauge the empathetic and sympathetic responses elicited by the proposed experience, a third variant of the original scale \cite{Escalas:2003} was created. This time, the statements refer to how they personally reacted to the experiment, and whether they felt as though they were embodying having a real visual impairment. By applying this to sighted people \textit{in loco}, after their experience takes place, and comparing these results to that of the pre-test, we expect to corroborate \textbf{H2}.
\end{itemize}


\subsection{Expert Feedback}
Prior to the beginning of the tests, a second visually impaired researcher participated in the experiment following the same protocols as the rest of the tests with sighted people would (as described in~\ref{experiment}) however, qualitative data was collected instead of the post-test ARS and ARE questionnaires. After the tests, the expert was interviewed regarding their thoughts about the research, and whether the sound interface created could be useful for BVI people in real scenarios. The participant provided valuable insights on how to shape the hypotheses present here, particularly noting that most academics believe the lived experience of disability cannot be equated by sighted people, so it is necessary to make it clear when addressing an actual feeling of empathy, as opposed to a self-reported one. Thus, when interpreting the results, we also ensured to avoid assuming that the empathetic and sympathetic responses were ground truth, but rather self-reported data prone to bias from misinformation, lack of experiences, or cognitive biases.

Additionally, they stated that it is critical to avoid simulating a visual impairment by solely making use of a blindfold, since this is believed to reduce the judged capabilities of people with disabilities, as it is only highlighting the initial challenges of becoming disabled. Based on that, we centered the experimental design around the assistive technology proposed instead, by incorporating relevant questions in the survey as well.

Regarding the sonified AR methods tested, the researcher added that it was important for audio guidance cues for specific objects not to overlap each other, making sure the user is following only one at a time, as to avoid confusion and minimize cognitive load. Along the trials, they also noted how easier it was to follow the methods with more practice, and that they learned to interpret the different sound attributes one by one, instead of handling multiple at the same time. When questioned about possible use cases, they suggested it would be likely more useful for indoor guidance, especially in unknown or complicated locations, rather than outdoors or in already familiar settings.

\section{Results}
\label{section:results}

Overall, 97 volunteers participated in the BVI perception and sighted pre-test survey collectively, out of which 20 were BVI (12 people totally blind and 7 with low vision, most having their current sight condition from birth, with the most recent being for the last 10 years). Although the subsample of BVI people is smaller, this is in line with prior work in the field~\cite{Lock:2020, Liu:2018, pare:2021, Coughlan:2020, Thevin:2020}, who also note the difficulty of recruiting this demographic group, oftentimes including blindfolded sighted users as well, in order to make comparisons between the groups and assess a greater population in the experiments. All survey respondents were in an age group between 18 and 80 (Mean = 35.23, SD = 13.22), with the sighted population between 18 and 80 (Mean = 32.60, SD = 11.32), and the BVI population between 27 and 74 (Mean = 46.75, SD = 14.24). Overall, 54 of the subjects identified themselves as female (55.67\%), 40 as male (41.23\%) and 3 as Non-Binary or Agender (3.09\%).

For the in-person experiment, 16 people volunteered to participate, out of which 1 had low vision, who was invited to participate and provide feedback as an expert in digital accessibility. The age range of participants was from 22 to 50 (Mean = 31.81, SD = 7.12), with 10 people identified themselves as female (62.5\%) and 6 as male (37.5\%). 8 people (50.00\%) declared to have at least some knowledge of musical theory, while 8 (50.00\%) claimed to have none. 11 people (68.75\%) said that they usually prefer listening to speech audio (e.g. podcasts, internet videos) at regular speed (1x), while 5 (31.25\%) claimed to prefer a faster-than-regular speed (1.25x or more). The experiment results do not include any volunteers who reported any hearing disabilities. No statistically significant difference was found in comparisons between their age, gender, musical literacy, or speech audio speed preference.


A Chi-Square test for normality was applied to all variables in the dataset in order to ascertain whether they were normally distributed. In order to verify their statistical significance (considering $\alpha = 0.05$), a Welch's unequal variances t-test with Bonferroni $p$ value adjustment was applied to all pairwise comparisons in which both subsets (BVI-Sighted and Pre-Post results) presented a normal distribution. For the remainder of the comparisons, in which at least one set was assumed to be not normally distributed, a Wilcoxon rank-sum test with Bonferroni $p$ value adjustment was used. The results for the specific attributes of the ARS and ARE scales are listed in Table \ref{resultsTable}, as well as the statistical significance test results of their comparisons.

\begin{table*}[t]
\caption{Pre and post experiment emotional scale results, including their mean (M), standard deviation (ST), and the statistical significance of their paired comparisons. \\ 
NND = Not Normally Distributed; NS = Not Statistically Significant}

\label{resultsTable}
\begin{tabular} {l|l|l|l|l|l}

\thead{\textbf{Emot.}\\ \textbf{Scale}}
& \thead{\textbf{BVI perception survey} \\ (N = 20)}
& \thead{BVI-Sighted\\Comparison}
& \thead{\textbf{Sighted pre-test} \\ (N = 77)}
& \thead{Pre-Post\\Comparison}
& \thead{\textbf{Sighted post-test} \\  (N = 15)}  \\
\hline\hline

ARS-1  & 
\thead{M = 2.95 SD = 1.19 \\ 1 \includegraphics[width=0.18\linewidth]{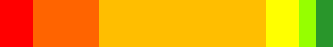} 7
}  
& \thead{$p \approx 0.18$  \\ \textbf{NS} (ttest) } 
& \thead{M = 3.5  ST = 1.78 \\ 1 \includegraphics[width=0.18\linewidth]{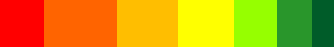} 7
}
& \thead{$p \approx 8.6e-3$  \\ \textbf{**} (ttest) }
&
\thead{M = 5.26  ST = 1.53 \\ 1 \includegraphics[width=0.18\linewidth]{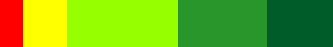} 7
}    \\ \hline

ARS-2 &
\thead{M = 2.65  ST = 1.34 \\ 1 \includegraphics[width=0.18\linewidth]{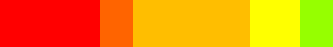} 7
}  
& \thead{$p \approx 0.018$ \\ \textbf{*} (ttest) } 
& \thead{M = 3.63  ST = 1.73 \\ 1 \includegraphics[width=0.18\linewidth]{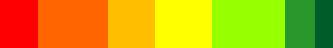} 7}  
& \thead{$p \approx 0.022$  \\ \textbf{*} (ttest) }
&  \thead{M = 4.93  ST = 1.53 \\ 1 \includegraphics[width=0.18\linewidth]{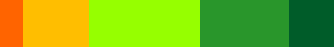} 7
}    \\ \hline

ARS-3 &
\thead{M = 3.45  ST = 1.50 \\ 1 \includegraphics[width=0.18\linewidth]{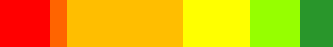} 7
} 
& \thead{$p \approx 2.4e-4$ \\ \textbf{***} (ttest)} 
& \thead{M = 4.89  ST = 1.51 \\ 1 \includegraphics[width=0.18\linewidth]{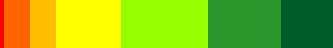} 7}   
& \thead{$p \approx 0.01$  \\ \textbf{*} (ttest) }
&  \thead{M = 5.40 ST = 1.35 \\ 1 \includegraphics[width=0.18\linewidth]{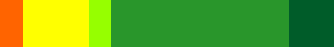} 7
}    \\ \hline

ARS-4 &
\thead{M = 2.95  ST = 1.66 \\ 1 \includegraphics[width=0.18\linewidth]{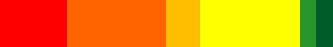} 7
}  
& \thead{$p \approx 2.3e-4$ \\ \textbf{***} (wilcox)} 
& \thead{M = 4.64 ST = 1.77 (\textbf{NND}) \\ 1 \includegraphics[width=0.18\linewidth]{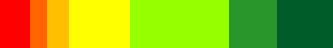} 7}   
& \thead{$p \approx 0.21$  \\ \textbf{NS} (wilcox) }
&  \thead{M = 5.13  ST = 1.64 \\ 1 \includegraphics[width=0.18\linewidth]{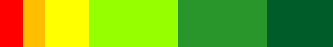} 7
}    \\ \hline

ARS-5 &
\thead{M = 3.05  ST = 1.50 \\ 1 \includegraphics[width=0.18\linewidth]{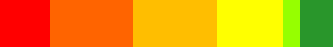} 7
}
& \thead{$p \approx 5.6e-3$ \\ \textbf{**} (ttest)} 
& \thead{M = 4.19  ST = 1.66 \\ 1 \includegraphics[width=0.18\linewidth]{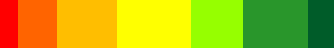} 7}
& \thead{$p \approx 4.5e-3$  \\ \textbf{**} (ttest) }
&  \thead{M = 5.73  ST = 1.03\\ 1 \includegraphics[width=0.18\linewidth]{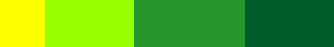} 7
}    \\ \hline\hline

\textbf{ARS} &
\thead{M = 3.01  ST = 1.44 \\ 1 \includegraphics[width=0.18\linewidth]{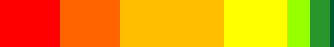} 7
} 
& \thead{$p \approx 1.8e-9$\\ \textbf{***} (ttest) }
& \thead{M = 4.17  ST = 1.77 \\ 1 \includegraphics[width=0.18\linewidth]{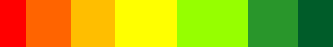} 7}  
& \thead{$p \approx 4.5e-7$  \\ \textbf{***} (wilcox) }
&  \thead{M = 5.29  ST = 1.42 \textbf{(NND)} \\ 1 \includegraphics[width=0.18\linewidth]{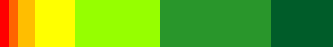} 7
}    \\ \hline\hline

ARE-1 &
\thead{M = 2.35  ST = 1.66\\ 1 \includegraphics[width=0.18\linewidth]{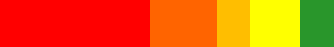} 7
}  
& \thead{$p \approx  0.037$ \\ \textbf{*} (ttest)}
& \thead{M = 3.31 ST = 1.89  \\ 1 \includegraphics[width=0.18\linewidth]{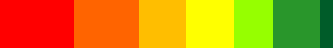} 7} 
& \thead{$p \approx 4.6e-3$  \\ \textbf{**} (ttest) }
&  \thead{M = 5.33  ST = 1.58  \\ 1 \includegraphics[width=0.18\linewidth]{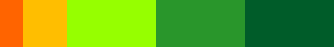} 7
}    \\ \hline

ARE-2 &
\thead{M = 1.80  ST = 1.43 (\textbf{NND}) \\ 1 \includegraphics[width=0.18\linewidth]{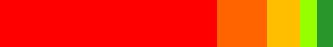} 7
} 
& \thead{$p \approx 0.023$ \\ \textbf{*} (wilcox)} 
& \thead{M = 2.73  ST = 1.86 (\textbf{NND}) \\ 1 \includegraphics[width=0.18\linewidth]{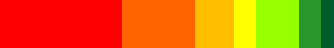} 7} 
& \thead{$p \approx 0.013$  \\ \textbf{*} (wilcox) }
&  \thead{M = 4.46  ST = 1.99 \\ 1 \includegraphics[width=0.18\linewidth]{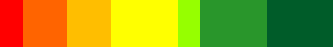} 7
}    \\ \hline

ARE-3 &
\thead{M = 1.9  ST = 1.41 (\textbf{NND}) \\ 1 \includegraphics[width=0.18\linewidth]{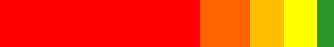} 7
} 
& \thead{$p \approx 0.032$ \\ \textbf{*} (wilcox) }
& \thead{M = 2.78  ST = 1.85 \\ 1 \includegraphics[width=0.18\linewidth]{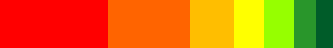} 7} 
& \thead{$p \approx 0.015$  \\ \textbf{*} (ttest) }
&  \thead{M = 4.80  ST = 1.89 \\ 1 \includegraphics[width=0.18\linewidth]{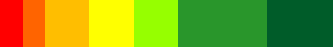} 7
}    \\ \hline

ARE-4 &
\thead{M = 2.45  ST = 1.63 \\ 1 \includegraphics[width=0.18\linewidth]{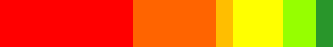} 7
} 
& \thead{$p \approx 0.095$ \\ \textbf{NS} (ttest) } 
& \thead{M = 3.18  ST =  1.80 \\ 1 \includegraphics[width=0.18\linewidth]{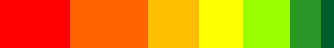} 7} 
& \thead{$p \approx 0.05$  \\ \textbf{NS} (ttest) }
&  \thead{M = 4.33  ST = 1.67  \\ 1 \includegraphics[width=0.18\linewidth]{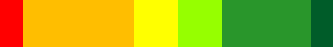} 7
}    \\ \hline

ARE-5 &
\thead{M = 2.25  ST = 2.02 (\textbf{NND}) \\ 1 \includegraphics[width=0.18\linewidth]{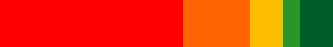} 7
}  
& \thead{$p \approx 0.038$ \\ \textbf{*} (wilcox) } 
& \thead{M = 2.96  ST = 1.75\\ 1 \includegraphics[width=0.18\linewidth]{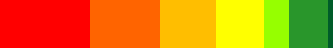} 7} 
& \thead{$p \approx 0.066$  \\ \textbf{NS} (ttest) }
&  \thead{M = 4.33  ST = 1.79  \\ 1 \includegraphics[width=0.18\linewidth]{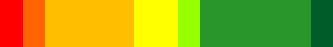} 7
}    \\ \hline\hline

\textbf{ARE} &
\thead{M = 2.15  ST = 1.63 (\textbf{NND}) \\ 1 \includegraphics[width=0.18\linewidth]{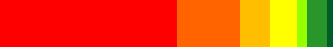} 7
}  
& \thead{$p \approx 3.6e-6$ \\ \textbf{***} (wilcox) }
& \thead{M = 2.99  ST = 1.83 (\textbf{NND}) \\ 1 \includegraphics[width=0.18\linewidth]{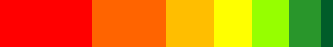} 7}  
& \thead{$p \approx 3.5e-7$  \\ \textbf{***} (wilcox) }
&  \thead{M = 4.65 ST = 1.78 \\ 1 \includegraphics[width=0.18\linewidth]{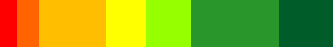} 7
}    

\end{tabular}
\end{table*}

\subsection {BVI-Sighted Comparison}
\label{comp1}
When comparing BVI people's perceived (BVI perception survey) to sighted people's self-reported (sighted pre-test) empathetic and sympathetic responses regarding their previous experiences with the other demographic, an increase of 16.57\% on the ARS scale ($p \approx 1.8e-9 $), and of 12.00\% on the ARE scale ($p \approx 3.6e-6 $) was noticed. This implies that people with visual impairment generally perceive sighted people to sympathize and empathize with their BVI condition significantly less than sighted people self-reported. The largest differences in each scale were observed in ARS-4 (24.14\% increase, $p \approx 2.3e-4 $), which addresses understanding the motivation of BVI people, and ARE-1 (13.71\% increase, $p \approx 0.037 $), which regards ``feeling as if the events were really happening to them". For full list of results refer to Table~\ref{resultsTable}.

\subsection {Sighted Pre-Post Comparison}
\label{comp2}
When comparing sighted people's self-reported pre-test to their self-reported post-test empathetic and sympathetic responses increase of 16.00\% on ARS ($p \approx 4.5e-7 $) and 23.71\% on ARE ($p \approx 3.5e-7 $), was noticed. This suggests that sighted people in general will feel significantly more sympathetic and empathetic to the BVI condition after participating in the experiment than they previously were. The largest differences in each scale were observed in ARS-1 (25.14\% increase, $p \approx 8.6e-3 $), which regards understanding the feelings of BVI people, and ARE-3 (28.85\% increase, $p \approx  0.01$), which addresses feeling as though they had a visual impairment. For full list of results refer to Table~\ref{resultsTable}.


\section{Discussion}
\label{section:discussion}
The current study aimed to explore sighted people's empathy with the BVI community, the BVI community's perception of said empathy, and the role of an embodiment of difference AR simulation of a total loss of vision on sighted people's empathy. An exploratory survey was also administered to investigate the perceived necessity of the development of new assistive technologies.

\subsection{Perceived and Self-Reported Empathy}
The results in \ref{comp1} support the~\textbf{H1} prediction that there would be a significant difference between sighted people's empathy with the BVI community and BVI people's perception of said empathy, as sighted people had a significantly higher score on both the ARS and ARE scales in comparison to BVI people. There are a number of interpretations that can be drawn from this. Firstly, it may be that sighted people believe themselves to be more empathetic than what they truly are. This may be due to social desirability bias, which is the tendency to choose responses that are more socially acceptable\cite{grimm2010social}. Although those that did not participate in the in-person experience were completely anonymous, this bias can be unconscious, meaning the participant would be unaware they were responding in a biased way. This difference between perceived and self-report empathy could also be due to sighted people not conveying this self-reported empathy strongly towards the BVI community, or their experiences do not involve in-person or direct communication with BVI people. An opposing explanation is that BVI people have had more negative experiences with sighted people on average, and due to this, they do not perceive the people from these experiences as having empathy towards the BVI community, which is then generalised to the wider sighted population.

In contrast with the original ARS and ARE scales, the pre-test questionnaire asked the sighted people of their broad experiences with the BVI community, without further detail. These experiences could have varied in both frequency and proximity across the sample, such that some people may have a BVI relative, but others could have interacted with media that features these people. With a further breakdown of these experiences, we may better understand how these do or not evoke empathy. 

\subsection{AR Experience and Empathy}
The results in \ref{comp2} support the \textbf{H2} prediction that sighted people's ARS and ARE scores would significantly increase after experiencing the BVI embodied experience in AR. This is consistent with the research that demonstrates embodiment of difference experiences increases empathy with people of different backgrounds and experiences~\cite{yee2011men,groom2009influence,yee2007proteus}, but more specifically disabilities~\cite{bacchus2019life, calepso2020, Guarese:Xchange, guareseEmpathy:2021}, which is the focus of this AR experience. The participants experienced this embodiment of difference due to being blindfolded and having to unfamiliarily rely heavily on senses other than sight. Also, they used a proposed assistive technology similar to what a BVI person may in the future have in their home to navigate their day-to-day tasks. This significant increase post-experiment is likely explainable by a substantial lack of previous experiences sighted people, in general, may have with the BVI community, with this experiment giving sighted people a direct and immersive insight into a BVI embodiment experience.
These results also demonstrate that an AR experience such as this, not only VR as previously researched~\cite{empathyLitReview,kandaurova2019effects,bujic2020empathy}, is also able to evoke empathy and is a valuable tool in providing embodiment of different experiences for social change.

Whilst social desirability bias may be an explanation for the discrepancy in self-reported and perceived empathy, it is also a potential limitation of the responses from the participants that completed the in-person experience. Due to the nature of the experiment, it was less anonymous and was conducted by a researcher of the study. Related, the demand effect is the tendency for respondents to respond in a way the researcher desires~\cite{zizzo2010experimenter}, which in the post-experiment tests, may have contributed to the significant increase. Although only the necessary information was communicated to the participant, they can make assumptions about the desired outcomes of the research.

Although the environment is set up to be a facsimile of a kitchen space, a wider variety of tasks for the experiment would have aided in a more accurate representation of a home, such as making a hot drink, retrieving cooking utensils, or washing the dishes. Similarly, as the HoloLens 2 and a number of other AR devices are not readily available to consumers, at the moment this technology may not be truly representative of the types of devices BVI people use to interact with their home. However, as technology advances and research such as this continues, it is hoped that these devices can become common in households. 

\section{Conclusions and future work}
\label{section:conclusion}

This work demonstrates an entertaining, challenging, and innovative way of promoting empathy with the BVI community through an interactive experience. Along with the feeling of touch on the real environment, and the use of voice commands for interaction, it was possible to create an experience in which a user without any impairments has a simulated total loss of vision, while also making use of a proposed assistive technology to be guided through simple object-placement tasks in a kitchen environment. Although simulations involving blindness in some sense already exist, the use of AR devices as assistive technology for micro-guidance has not been widely studied in the same context.

Building on the findings regarding a difference in perceived and self-reported empathy from sighted people, it would be beneficial to gather further data on the opinions and perceptions of sighted people on the current accessibility of everyday interactions, and to compare this to the BVI community's experiences. These comparisons would provide valuable insight to include when creating authentic embodiment of difference experiences and to understand the extent of this difference in perception. 


Future work includes the use of the performance data gathered from the participants into a preliminary study on the usability of similar types of sonified assistive technology methods. After optimizing said methods, we aim to test their performances while being used by BVI people, which we hypothesize are likely to have behaviors in place facilitating themselves in environments such as this. By observing their use, we expect to improve our methods considering their needs, developing new assistive technologies that can be used on real-life scenarios, besides the simulated environment. Regarding the survey results regarding the need for new assistive technologies, our preliminary findings suggest there is a broad perceived need of these methodologies. Thus, we also plan to explore in more depth the specific areas in need of these assistive technologies by conducting a bigger survey to the visually impaired community. Beyond that, we aim to extend the experiment towards other types of blindness. With the use of eye-tracking features, it is possible to add visual artifacts that can mimic other vision disorders, such as severe retinopathy or corneopathy, which can be very disabling, showing how visually impairments are not all the same.

\acknowledgments{
The authors wish to acknowledge the contribution of our late colleague and co-author Dr. Ron van Schyndel. Ron passed away in late December, during the preparation of this work. He will be sorely missed by his colleagues and students.

Dr. Ron van Schyndel was a Senior Lecturer at the School of Computing Technologies at RMIT University. He was an active researcher in the domain of digital watermarking for more than two decades, and is co-author to some of the most cited papers in the field. His research interests beyond digital watermarking, included signal, image and vision processing, as well as mobile navigation for people who are blind and visually impaired, being visually impaired himself. He published over 50 papers in these fields in top conferences and journals. He obtained his Ph.D. from Monash University on the nascent topic of digital watermarking, and has obtained many industry grants on watermarking applications, including image, audio, and video on desktop servers and mobile devices. Prior to his academic career, he worked for organisations such as Toshiba Medical in medical imagery, DeBeers Mining in satellite imagery, as programmer for several mobile telcos, and as scientific programmer for Monash University. He chaired the local IEEE Computer Society chapter from 2006 to 2010, making it the most active chapter in the Victorian section at the time.

Beyond that, we would like to thank all experiment participants for their time, the Australian Technology Network of Universities for providing funding through the ATN-Latin American PhD Scholarship, and RMIT for the Research Stipend Scholarship.

}


\bibliographystyle{abbrv-doi}

\bibliography{template}

\end{document}